\documentclass[amsmath,amssymb,aps,reprint,
prb]{revtex4-1}

\usepackage{chemformula} 
\usepackage[T1]{fontenc} 

\usepackage{graphicx}

\begin{document} 

\author{Sofiane Haffouz}
\email{sofiane.haffouz@nrc.ca}
\author{Dan Dalacu}
\author{Philip J. Poole}
\author{Jean Lapointe}
\author{Daniel Poitras}
\author{Khaled Mnaymneh}
\author{Xiaohua Wu}
\author{Marek Korkusinski}
\author{Robin L. Williams}

\affiliation{National Research Council of Canada, Ottawa, Ontario, Canada, K1A 0R6.}

\title{Bright single InAsP quantum dots at telecom wavelengths in position-controlled InP nanowires: the role of the photonic waveguide}







\begin{abstract}
We report on the site-selected growth of bright single InAsP quantum dots embedded within InP photonic nanowire waveguides emitting at telecom wavelengths. We demonstrate a dramatic dependence of the emission rate on both the emission wavelength and the nanowire diameter. With an appropriately designed waveguide, tailored to the emission wavelength of the dot, an increase in count rate by nearly two orders of magnitude (0.4\,kcps to 35\,kcps) is obtained for quantum dots emitting in the telecom O-band. Using emission-wavelength-optimised waveguides, we demonstrate bright, narrow linewidth emission from single InAsP quantum dots with an unprecedented tuning range from 880\,nm to 1550\,nm. These results pave the way towards efficient single photon sources at telecom wavelengths using deterministically grown InAsP/InP nanowire quantum dots. 

\end{abstract}

\maketitle 
\newpage
Non-classical light sources that can produce streams of correlated on-demand photons are a central building block for optics-based quantum information technologies. In recent years, there has been a major drive to demonstrate single photon sources for quantum key distribution (QKD) to provide secure communications over long distances.  Sources for QKD need to generate single photons with negligible probability of multiphoton emission and, for certain protocols, these photons need to be indistinguishable\cite{Lo_PRL2012,Lo_Science1999}. Ideally, the sources should produce single photons on-demand at high repetition rates which can be efficiently collected by an external optical system\cite{Aharonovich_NP2016}. Importantly, for fibre-based long-haul communications, the sources should emit photons with wavelengths in the telecom windows around $1.3\mu$m and $1.5\mu$m where fibre losses are minimised.

The InAs/InP quantum dot material system has been shown to emit efficiently at telecom wavelengths\cite{Poole_JVST2001,Dalacu_PRB2010}. Single photon emission at $1.5\mu$m was demonstrated in 2005\cite{Miyazawa_JJAP2005} and, more recently, bright single photon emission was demonstrated at $1.3\mu$m and $1.5\mu$m using photonic crystal cavities\cite{Kim_Optica2016,Birowosuto_SR2012}. These devices were based on randomly nucleated quantum dots, although site-selected dots\cite{Dalacu_PRB2010,Dalacu_NL2012}, nucleated at specified positions on the substrate, are a preferred growth mode. Site selection opens up the possibility for deterministic integration of telecom quantum dots with photonic structures aimed at, for example, increasing device efficiency\cite{Lodahl_RMP2015}.

The photonic nanowire approach to fabricating efficient quantum light sources has shown great promise, with demonstrated collection efficiencies of 72\% \cite{Claudon_NP2010}. Bottom-up based nanowire devices\cite{Borgstrom_NL2005} are of particular interest since they can naturally contain one and only one dot per device and they are readily grown using site-selective techniques\cite{Heinrich_APL2010,Dorenbos_APL2010,Holmes_NL2014}. Site-selected InAsP/InP nanowire quantum dot sources grown using a combined selective-area and vapour-liquid-solid (VLS) epitaxy approach\cite{Dalacu_NT2009} have demonstrated high collection efficiencies of 43\% (corresponding to an 86\% coupling into the fundamental HE$_{11}$ nanowire waveguide mode), and near transform-limited linewidths of 4\,$\mu$eV\cite{Reimer_PRB2016}. Such sources can generate single photons with negligible multiphoton probabilities ($g_2(0) < 0.005$\cite{Dalacu_NL2012}) and polarization entangled photon pairs (via the biexciton-exciton cascade) with fidelities to the maximally entangled state exceeding 80\%\cite{Jons_SR2017,Versteegh_NC2014,Huber_NL2014}. These sources utilize InAs$_x$P$_{1-x}$ quantum dots with $x\sim 30$\% and operate at wavelengths of $\lambda \sim 950$\,nm.

It is well known that the diameter of the photonic nanowire will determine the spontaneous emission rate of the quantum dot by affecting the available optical density of states, i.e. by dictating the overlap of the waveguide mode (e.g. HE$_{11}$) with the dipole field of the emitter\cite{Friedler_OE2009}. Previous studies have shown an order of magnitude reduction in transition lifetimes\cite{Bulgarini_APL2012} as well as more modest increases in brightness\cite{Jeannin_PRApp2017} by optimising the nanowire diameter. These studies tuned the photonic nanowire diameter and used dots emitting at a fixed energy. In this study we first tune the quantum dot towards telecom wavelengths for a given nanowire diameter, producing  a dramatic drop in emission intensity with increasing wavelength, and we then illustrate the recovery of the lost intensity by increasing the nanowire diameter at this longer wavelength. Using waveguides tailored to the emission wavelength of the dot, we demonstrate bright, narrow linewidth single dot emission tunable from $\lambda = 880$\,nm up to $\lambda = 1550$\,nm.

The devices used in this study consist of a nanowire core with one or two embedded InAsP quantum dots that are clad with an InP shell (see Methods). The diameter of the core and of the dots is determined by the size of the gold catalyst particle (nominally 18\,nm for these studies), which is itself defined by electron-beam (e-beam) lithography. The diameter of the cladding, which defines the waveguide, is determined by the size of a circular opening in a SiO$_2$ mask defined by the time of a wet etch step. The top of the cladding is tapered to improve coupling to the external optical system\cite{Gregersen_OL2008}. Details of the device growth are given in Refs. \citenum{Dalacu_NT2009,Dalacu_NL2012}.

\begin{figure}
\begin{center}
\includegraphics*[width=8cm,clip=true]{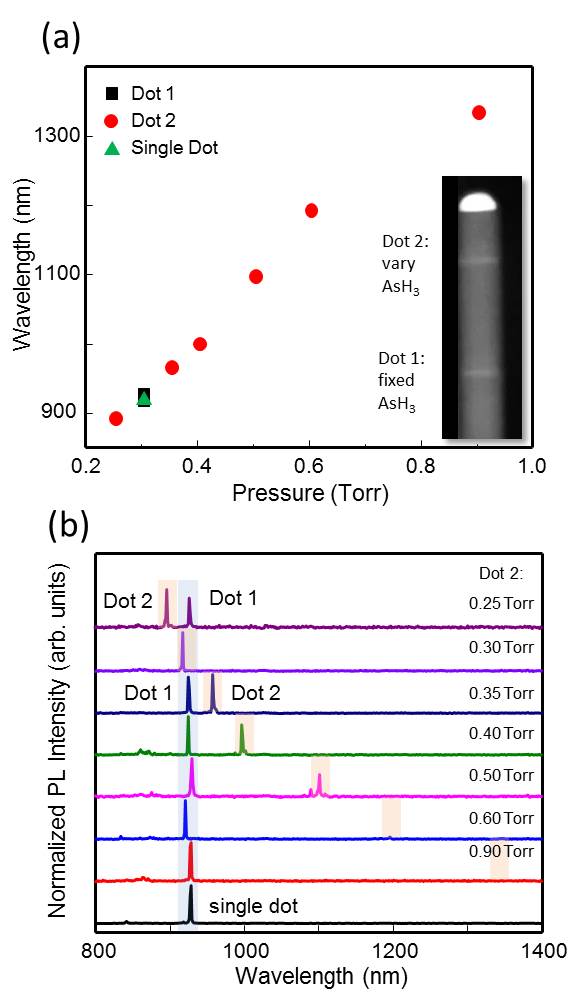}
\end{center}
\caption{(a) Emission wavelength and (b) PL spectra of the double dot system versus AsH$_3$ flux used for Dot 2. Inset in (a) shows a transmission electron microscopy image of a double dot nanowire core with a 18\,nm diameter.}\label{fig1}
\end{figure}

To highlight the role of the waveguide design in determining the emission properties of the quantum dot emitters, we grew double quantum dot samples (i.e. two dots in the same photonic nanowire waveguide) and varied the emission wavelength of one of the dots. Seven samples were grown with the two quantum dots incorporated in the nanowire core separated sufficiently to avoid any electronic coupling (see inset, Fig. 1(a)). The InAs$_x$P$_{1-x}$ quantum dot emission energy was tuned by controlling the dot composition, $x$, through variation of the arsine flux. For the first dot (Dot 1) the growth conditions were the same for all samples and targeted an emission wavelength in the 900 - 950\,nm range (AsH$_3$ pressure = 0.3\,Torr). For the second dot (Dot 2), the AsH$_3$ flow was varied for each sample (0.25 to 0.9\,Torr). The growth time for each dot was 3.5 seconds and the finished waveguide diameter was $D =200$\,nm. Fig. 1(b) shows the photoluminescence (PL) spectra from the seven double dot nanowires as well as from a single dot reference sample, all measured at an excitation power $P = 0.25P_{sat}$ where $P_{sat}$ is the excitation power required to saturate the ground state transition in the reference sample. 

Dot 1 is seen to emit at $\lambda \sim 920$nm, with little variation from sample to sample (less than $\pm 5$\,nm), highlighting the reproducibility of the growth process. The emission wavelength of Dot 2 varies from $\lambda \sim 900$\,nm for 0.25\,Torr to $\lambda \sim 1200$\,nm for 0.6\,Torr. The shift to longer wavelengths is accompanied by a dramatic decrease in the emission intensity, so much so that emission from Dot 2 for 0.9\,Torr is completely absent. For this dot, an emission wavelength of $\lambda = 1337$\,nm was determined by pumping at $P = P_{sat}$. 

\begin{figure}
\begin{center}
\includegraphics*[width=9cm,clip=true]{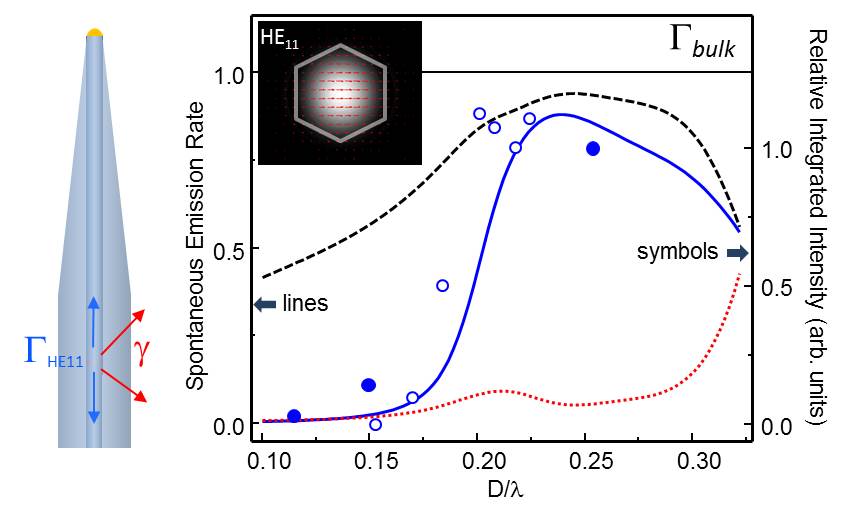}
\end{center}
\caption{Calculated spontaneous emission rate into the fundamental HE$_{11}$ nanowire waveguide mode ($\Gamma_{HE11}$ - blue solid line) and into leaky modes ( $\gamma$ - red dotted line) of an artificial atom placed on the axis of an InP nanowire as a function of the normalized wire diameter ($D/\lambda$). Black dashed line is $\beta = \Gamma_{HE11}/(\Gamma_{HE11} + \gamma)$. Open and filled circles are measured integrated PL intensities at saturation from Dot 2 of the double dot sample and the a-Si coated sample, respectively. Inset shows the calculated electric field intensity of one of the two orthogonally polarized HE$_{11}$ modes.}\label{fig2}
\end{figure}

To evaluate the emission properties of the InAsP/InP quantum dot nanowires, we have calculated the spontaneous emission rate into the fundamental HE$_{11}$ nanowire waveguide mode ($\Gamma_{HE11}$) for an electric dipole placed on the axis of the nanowire and oriented such that the dipole and nanowire axes are orthogonal. Emission rate data is presented in Fig. 2, normalized to the calculated spontaneous emission rate in bulk InP, $\Gamma_{bulk}$, as a function of the normalized wire diameter, $D/\lambda$, as in Ref. \citenum{Friedler_OE2009}. A normalized spontaneous emission rate into HE$_{11}$ of close to $\Gamma_{HE11} = 0.9$ can be achieved using an optimum waveguide diameter $D/\lambda \sim 0.23$. $\Gamma_{HE11}$ drops rapidly for smaller values of $D/\lambda$. For a dot emitting at $\lambda  = 1360$\,nm for example, the optimal diameter is $D \sim 310$\,nm. Using a standard nanowire waveguide optimised for $\lambda  = 950$\,nm, with a diameter of $D=200$\,nm, corresponds to $D/\lambda \sim 0.15$ at $\lambda = 1360$\,nm and results in an emission rate that is reduced by a factor of 35 ($\Gamma_{HE11} = 0.024$). This reduction in spontaneous emission rate has two major effects; the saturated intensity of any given transition will be reduced as the radiative lifetime is increased \cite{Bulgarini_APL2012} and any competing non-radiative decay channels become more effective. An additional effect is a reduction in the fraction of photons coupled into the HE$_{11}$ mode, $\beta$, versus all other radiation modes where $\beta = \Gamma_{HE11}/(\Gamma_{HE11} + \gamma)$ and $\gamma$ is the emission rate into leaky modes. The value of $\beta$ drops from 93\% to 51\% as $D/\lambda$ changes from 0.23 to $0.15$. 

The combination of these effects can result in a dramatic decrease in the measured emission intensity when the photonic nanowire diameter is not optimized for the emission wavelength of interest. This decrease is observed directly in the measured spectra of Fig. 1 and is included in Fig. 2 (open circles)  as the integrated intensity from Dot 2 normalized to that of Dot 1.  This data shows clear qualitative agreement with the predicted behaviour as a function of $D/\lambda$. 

\begin{figure}
\begin{center}
 \includegraphics*[width=9cm,clip=true]{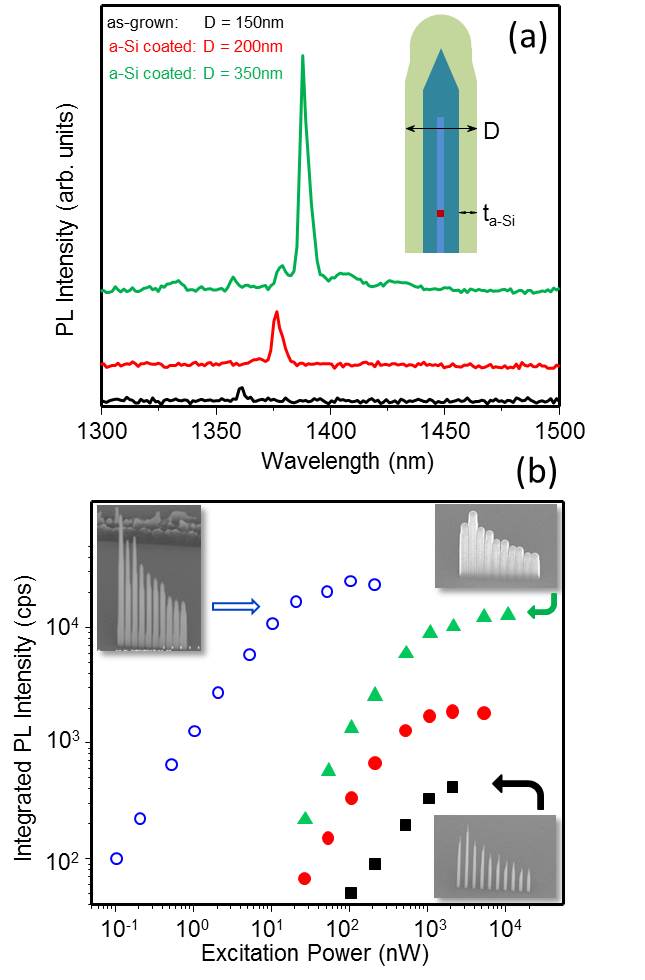}
\end{center}
\caption{(a) PL spectra of the same nanowire as-deposited and coated with a-Si. (b) Corresponding integrated PL intensities of the ground state emission: as-deposited: $t_{a-Si} = 0$\,nm,  $D=150$\,nm (black squares), first coat: $t_{a-Si} = 25$\,nm, $D=200$\,nm (red circles), and second coat: $t_{a-Si} = 75$\,nm, $D=350$\,nm (green triangles).  Blue open circles correspond to the ground state emission from a quantum dot in an all InP photonic nanowire having an as-deposited $D=350$\,nm. The two right-hand insets show scanning electron microscopy (SEM) images of the same linear array of nanowires (400\,nm pitch) as grown (top inset) and after the second a-Si coating (bottom inset). The left hand inset shows an SEM image of a linear array of as-grown nanowires with $D=350$\,nm.}\label{fig3}
\end{figure}

To demonstrate that the observed drop in PL intensity with increasing dot emission wavelength is influenced by the waveguide design, we use a single dot nanowire and increase the diameter of the waveguide incrementally by depositing additional cladding material. The single dot was incorporated in a photonic nanowire with diameter $D = 150$\,nm. The growth conditions were chosen to target an emission wavelength of $\lambda \sim 1350$\,nm (AsH$_3$ pressure = 1.2\,Torr, growth time = 3.5\,s). Sequential coatings of amorphous silicon (a-Si) having a thickness $t_{a-Si}$ were then sputtered on the nanowire sidewall to incrementally increase the cladding diameter, see inset Fig. 3(a). InP and a-Si have similar indices of refraction at these wavelengths (3.2 and 3.4 respectively) and we assume  that the a-Si cladding will simply increase the effective index of the waveguide and not affect the mode structure dramatically. Fig. 3 shows the PL from a single dot as grown ($D = 150$\,nm), after deposition of $t_{a-Si} = 25$\,nm ($D = 200$\,nm) and after $t_{a-Si} = 75$\,nm ($D = 350$\,nm). With each coating, the PL intensity at saturation increases by close to an order of magnitude. The integrated intensities at saturation, normalized to that from the nanowire after the $t_{a-Si} = 75$\,nm coating, are included in Fig. 2 (filled circles). This data also shows clear qualitative agreement with the predicted behaviour as a function of $D/\lambda$.  We note that we also observe a slight red-shift of the emission wavelength with each a-Si coating. The sputtered a-Si on InP is compressively strained\cite{Windischmann_NCS1986} and hence applies a tensile strain on the nanowire resulting in a red-shift of the dot emission that depends on the thickness of a-Si deposited \cite{Bavinck_NL2012}.

We look next at growing larger diameter waveguides tailored to the dot emission wavelength $in$ $situ$, rather than using an a-Si coating. The growth mode employed allows one to independently control the diameter of the nanowire core and that of the cladding\cite{Dalacu_NL2012}. The upper panel in Fig. 4 shows schematically three of the processing steps involved in the preparation of the growth substrate. The second step defines the opening in the SiO$_2$ mask that permits the selective-area growth of the cladding. The size of this opening is determined by a hydrofluoric wet-etch which, in turn, determines the diameter of the clad nanowire. For example, by increasing the hole diameter from $D_{hole} = 100$\,nm (12\,s etch) to $D_{hole} = 300$\,nm (24\,s etch) the diameter of the clad nanowire increases from $D = 220$\,nm to $D = 340$\,nm, as shown in the middle and lower panels, respectively, of Fig. 4. 

\begin{figure}
\begin{center}
\includegraphics*[width=8cm,clip=true]{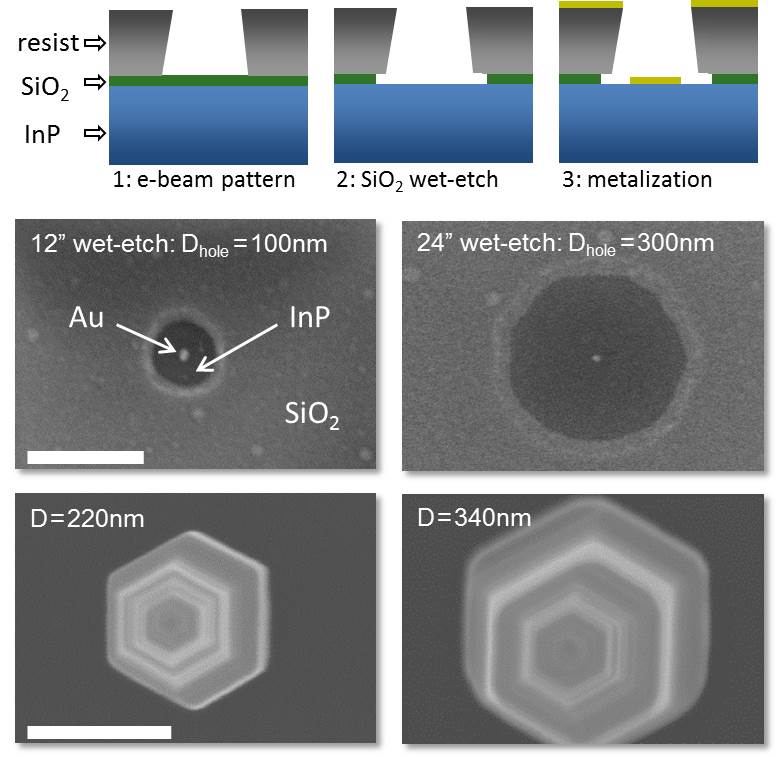}
\end{center}
\caption{Upper panel: processing steps for the preparation of the growth substrate. Grey is e-beam resist, green is SiO$_2$, blue is the InP substrate, and yellow is gold. Middle panel: SiO$_2$ hole opening for different etch times. Lower panel: top-view SEM of nanowires grown in the different size openings. Scale bars are 200\,nm.
}\label{fig4}
\end{figure}

We have used this control of the cladding diameter to design a waveguide tailored to the dot emission wavelength (e.g. $D/\lambda = 0.23$). We targeted an emission wavelength in the $\lambda =1350$\,nm range and a cladding diameter of $D= 350$\,nm. To reach this wavelength we tuned both the composition (increased AsH$_3$ flow) and  degree of confinement (longer growth time) of the InAsP dot. In particular, the AsH$_3$ flow and growth time were 3\,sccm and 9\,s, respectively, compared to 2\,sccm and 3\,s for dots emitting at $\lambda = 950$\,nm\cite{Dalacu_NL2012}. Figure 5(a) shows the power dependent PL spectra of the device. The ground state emission (X$^-$) is at $\lambda = 1342$\,nm with a measured linewidth of $\sim 150\mu$eV, limited by the resolution of the spectrometer. The waveguide diameter is $D = 340$\,nm. The power-dependent integrated PL intensities are included in Fig. 3 for comparison with the a-Si coated structures. There is a striking difference in the power required to saturate the ground state transition between this sample and the a-Si coated sample of similar diameter. We speculate that this is related to the difference in absorption of the excitation laser and subsequent diffusion of carriers to the quantum dot between the two structures. The a-Si and its interface with the InP are likely trapping the majority of the optically generated carriers resulting in an excitation power required for saturation that is close to two orders of magnitude larger.

\begin{figure}
\begin{center}
 \includegraphics*[width=8cm,clip=true]{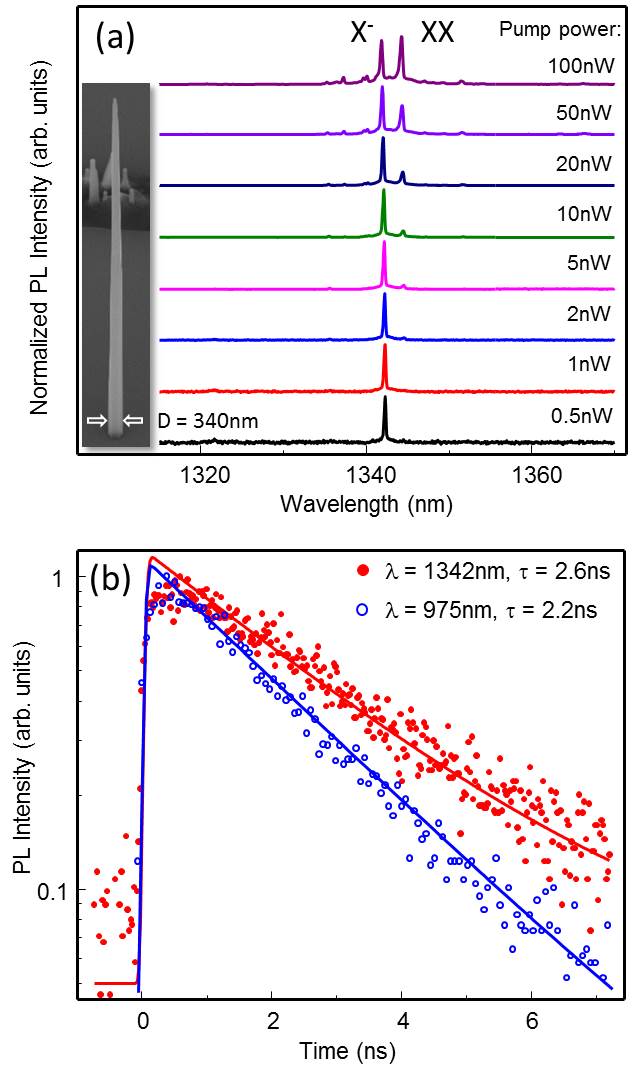}
\end{center}
\caption{(a) Power dependent spectra for a dot with a charged exciton ground-state emission  $\lambda = 1342$\,nm in a $D=340$\,nm waveguide shown in inset (e.g. $D/\lambda = 0.25$). (b) Decay curves for two nanowire dots with the same normalized diameter but different ground-state emission wavelengths ($\lambda=1342$\,nm: red filled circles, $\lambda=975$\,nm: blue empty circles).
}\label{fig5}
\end{figure}

The second difference concerns the count rate at saturation. The emission rate from the InP clad nanowire with $D=340$\,nm is almost two orders of magnitude higher than the device with $D=150$\,nm (35\,kcps compared to 0.4\,kcps). The emission rate for the $D=340$\,nm InP clad nanowire is also two times larger than the comparable a-Si clad nanowire, even though $D/\lambda$ values are similar. The a-Si coating removes the taper present in the as-grown devices and one can expect a strong reduction in collection efficiency from both back reflections at the nanowire tip and from increased divergence of the emitted light\cite{Gregersen_OL2008,Friedler_OE2009}. Also, a small amount of NIR absorption can be expected in unpassivated a-Si (see Methods). 
 
From a measurement of the decay of the X$^-$ transition we obtain a lifetime of  $\tau = 2.6$\,ns, similar to the lifetime of a nanowire dot emitting at $\lambda = 975$\,nm with a similar normalized wire diameter. Both decay curves are shown in Fig. 5(b). These short lifetimes are consistent with an uninhibited emission process ( $\tau \sim \tau_{bulk}$), see Bulgarini $et$ $al$\cite{Bulgarini_APL2012}. Given the similar lifetimes, one would expect similar count rates from the sources at $\lambda = 975$\,nm and $\lambda = 1342$\,nm, assuming similar collection efficiencies (i.e. no difference in the far-field emission profile). The count rates at saturation from the two sources, taking account of the wavelength-dependent spectrometer efficiency, $\eta$, are 275\,kcps at $\lambda = 1342$\,nm (measured 35\,kcps, $\eta_{1310nm} = 12.7$\%) and 5900\,kcps at $\lambda=975$\,nm (measured 135\,kcps, $\eta_{980nm} = 2.3$\%). 

The power radiated from an electric dipole is expected to scale as $\omega^4$, corresponding to a count rate scaling as $\omega^3$. Such behaviour does not account for the large emission intensity changes observed here and we are forced to consider the structural changes to the quantum dot that were used to shift the emission wavelength from $\lambda =975$\,nm to $\lambda = 1342$\,nm. The AsH$_3$ flow was increased by 33\% and the growth time was tripled. From TEM analysis of nanowire cores, the longer quantum dot growth time increases the dot thickness from $h \sim 3$\,nm to $h \sim 7$\,nm. This corresponds to a much higher aspect ratio, $h/D_{dot} =7/18$, compared to typical self-assembled Stranski-Krastanow (SK) quantum dots. In SK dots, the strong confinement in the growth direction together with the resulting strain profile leads to a splitting of the heavy-hole (HH) and light hole (LH) valence levels with a HH ground state\cite{Korkusinski_PRB2013}. This produces two bright excitons polarized perpendicular to the growth direction. Atomistic tight-binding calculations of similar InAs/InP quantum dot structures\cite{Niquet_PRB2008, Zielinski_PRB2013} have shown a transition from a compressive to tensile biaxial strain profile and a corresponding transition to a LH ground state with increasing quantum dot aspect ratio. The excitons associated with a LH ground state will be polarized along the nanowire growth axis. LH excitons have previously been observed in nanowire quantum dots\cite{Jeannin_PRB2017}. 

For intermediate quantum dot aspect ratios ($h/D_{dot} = 8/9.6$) the ground state is expected to be a linear combination of HH and LH levels\cite{Zielinski_PRB2013}, producing a mixture of both perpendicular and parallel (with respect to the nanowire axis) polarized excitons. The latter is not expected to couple to the HE$_{11}$ mode\cite{Jeannin_PRApp2017} and the resulting radial emission would be collected with very low efficiency by the external optics. Considering the aspect ratio of the quantum dot in our device, it is likely that a significant fraction of the recombination is via the LH state, resulting in unguided photons and contributing to the observed drop in count rates.

Finally, we show that with appropriately designed waveguide structures targeting $D/\lambda \sim 0.23$, the InAsP/InP nanowire quantum dot system can cover a large part of NIR spectral range. Fig. 6 shows PL spectra of single InAsP/InP quantum dot nanowires grown under different growth conditions. By adjusting the arsenic flow and dot thickness as well as the dot diameter\cite{Dalacu_APL2011}, this single material system can be tuned from $\lambda = 880$\,nm to $\lambda = 1550$\,nm. Each peak is resolution-limited and corresponds to a charged exciton (X$^-$) ground state. We note that the drop in the emission rate observed when going from $\lambda = 975$\,nm to $\lambda = 1342$\,nm continues as the wavelength is tuned further and can be seen from the reduced single-to-noise in the normalized spectra for $\lambda > 1400$\,nm. The saturation count rate at the first lens for emission at $\lambda = 1532$\,nm is down to 19\,kcps.

\begin{figure}
\begin{center}
\includegraphics*[width=8cm,clip=true]{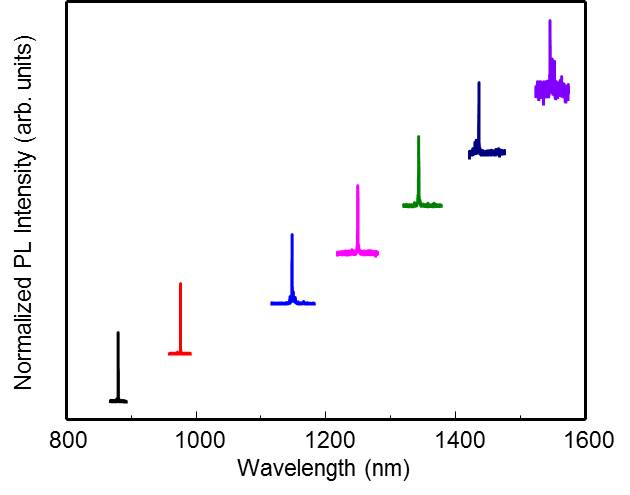}
\end{center}
\caption{Normalized PL spectra of single InAsP/InP nanowire quantum dot emitters in tailored photonic waveguides measured at an excitation power of 0.1\,P$_{sat}$.}\label{fig6}
\end{figure}

In conclusion, we have identified the design of the photonic nanowire waveguide to be crucial in obtaining bright emission from InAsP/InP nanowire quantum dots emitting at telecom wavelengths.  Using appropriately designed waveguides, we have demonstrated bright, narrow linewidth, single quantum dot emission from a single material system with an unprecedented tuning range spanning 880\,nm to 1550\,nm. We observe a larger than expected decrease in brightness with increasing wavelength in these optimised structures and suggest that an increasing radial component of the quantum dot dipole due to valence band mixing between light- and heavy-hole levels  may be a contributing factor.

\subsubsection{Methods}

Nanowire Growth: The growth of the InAsP/InP nanowire devices is carried using chemical beam epitaxy on Fe-doped InP (111)B substrates with trimethylindium and pre-cracked phosphine (PH$_3$) and arsine (AsH$_3$) as precursors of indium, phosphorus and arsenic, respectively. Two growth systems were used, one in which the precursor flux is controlled directly using a mass flow controller, and another where the flux was set by a pressure control system. The growth temperature is controlled by band-edge thermometry and typically is in the range 420 - 435$^{\circ}$C. The nanowires are grown using a combined selective-area and VLS process with Au catalyst particles (see Refs. \citenum{Dalacu_NT2009,Dalacu_NL2012} for details). Briefly, we first grow a nanowire core using a growth temperature and PH$_3$ flow that are tuned to obtain only axial growth via the VLS process ($420-435^{\circ}$C  and 2\,sccm, respectively). The quantum dot is incorporated in the core a few hundred nanometers from the base by switching AsH$_3$ in for PH$_3$. To grow the InP shell, the PH$_3$ flow is tripled in order to trigger radial growth and turn off the axial growth.

Amorphous silicon deposition: Ion-beam sputtering (Spector, Veeco) was used to deposit hydogen-free a-Si at room temperature. The measured extinction coefficent of the films was $k \sim 0.01$ for wavelengths $\lambda = 1300-1400$\,nm. 

Optical Spectroscopy: Optical measurements on individual nanowires were performed with the wires still attached to the (111)B InP substrate. The measurements were done at 4.2\,K in a continuous flow helium cryostat using non-resonant, above bandgap excitation through a 50X microscope objective (N.A. = 0.42) with a $\sim 2$\,$\mu$m spot size. The PL was collected through the same microscope objective, dispersed using a 0.320\,m grating spectrometer and detected using a liquid-nitrogen cooled InGaAs diode array. Lifetime measurements were performed with pulsed excitation from a diode laser at 670\,nm using 100\,ps pulses at a repetition rate of 80\,MHz and the emitted photons were detected by a superconducting nanowire single photon detector with a 40\,ps timing jitter.

\bibliographystyle{Prsty}

\providecommand{\latin}[1]{#1}
\makeatletter
\providecommand{\doi}
  {\begingroup\let\do\@makeother\dospecials
  \catcode`\{=1 \catcode`\}=2\doi@aux}
\providecommand{\doi@aux}[1]{\endgroup\texttt{#1}}
\makeatother
\providecommand*\mcitethebibliography{\thebibliography}
\csname @ifundefined\endcsname{endmcitethebibliography}
  {\let\endmcitethebibliography\endthebibliography}{}

\end{document}